\documentclass[a4paper,11pt]{article}
\pdfoutput=1

\usepackage{jcappub}

\usepackage[T1]{fontenc}

\usepackage{amsmath}
\usepackage{amssymb}
\usepackage{amsfonts}
\usepackage{bm}
\usepackage{verbatim}
\usepackage{tablefootnote}
\usepackage{enumerate}
\usepackage{afterpage}
\usepackage{xcolor}
\usepackage{natbib, hyperref}

\usepackage{graphicx}
\usepackage{tabularx}
\usepackage{booktabs}
\usepackage{pifont}

\newcommand{\orcid}[1]{\href{https://orcid.org/#1}{\,\includegraphics[width=8px]{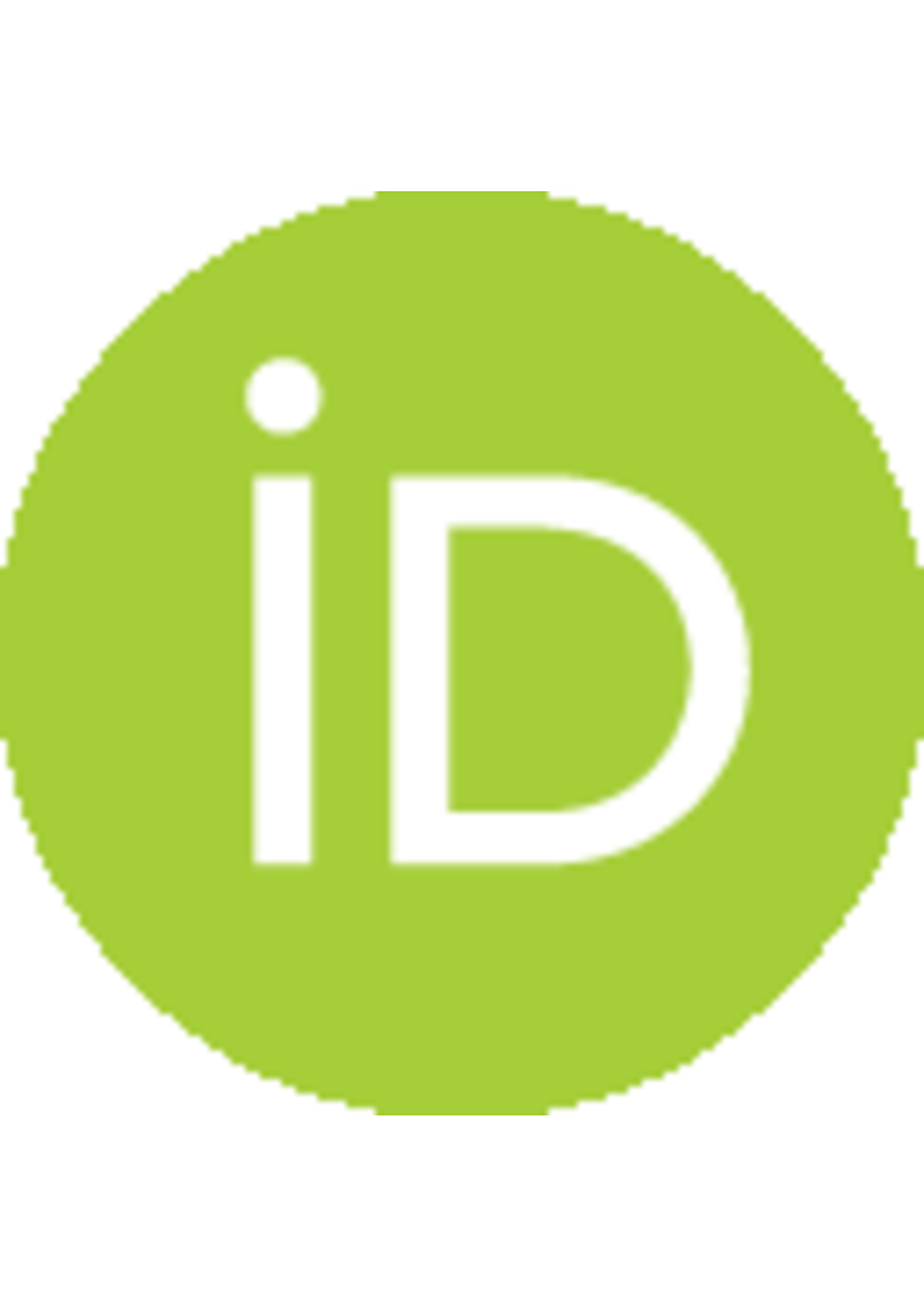}}}

\usepackage{soul}
\usepackage{xcolor}
\setstcolor{blue}

\usepackage{tikz}
\newcommand{\mathst}[1]{
  \tikz[baseline=(X.base)]{
    \node[inner sep=0pt, outer sep=0pt] (X) {$#1$};
    \draw[blue, line width=0.5pt] (X.west) -- (X.east);
}
}
\usepackage{etoolbox}
\robustify{\mathst}

\title{\boldmath  Calibration-independent consistency test of BAO and SNIa data: update}

\author{Bikash R. Dinda$^{a}$\orcid{0000-0001-5432-667X},
Roy Maartens$^{a,b}$\orcid{0000-0001-9050-5894},
Chris Clarkson$^{c,a}$\orcid{0009-0002-9207-5239}}

\affiliation[a]{Department of Physics \& Astronomy, University of the Western Cape, Cape Town 7535, South Africa}

\affiliation[b]{National Institute for Theoretical \& Computational Science, Cape Town 7535, South Africa}

\affiliation[c]{Department of Physics \& Astronomy, Queen Mary University of London, London E1 4NS, United Kingdom}

\emailAdd{bikashrdinda@gmail.com}
\emailAdd{rmaartens@uwc.ac.za}
\emailAdd{chris.clarkson@qmul.ac.uk}

\abstract{
In a recent paper \href{https://arxiv.org/pdf/2509.19899}{2509.19899}, we presented a new method to test the consistency between uncalibrated BAO and SNIa data through a common parameter, the Alcock–Paczynski variable. Using  Gaussian Processes, we can determine if various datasets are consistent, independently of dark energy or modified gravity models, and of the sound horizon and SNIa peak magnitude. We found that the DES-Y5 SNIa data showed non-negligible tension with other datasets. However, the recent update DES-Dovekie removes this tension. We find that all uncalibrated data from DESI DR2 BAO  and three SNIa datasets, Union3, Pantheon+, and DES-Dovekie, are consistent with each other within $\sim\! 1 \sigma$.
}

\notoc

\begin{document}
\maketitle
\flushbottom

\section{Introduction}

The Dark Energy Spectroscopic Instrument (DESI) data on baryon acoustic oscillations (BAO)  \citep{DESI:2024mwx,DESI:2025zgx}, show evidence of dynamical dark energy, i.e., the equation of state of dark energy $w\neq-1$. Using the $w_0w_a$ parametrization, the $\Lambda$CDM model is $3.1\sigma$, $3.8\sigma$, $2.8\sigma$ and $4.2\sigma$ away from the mean values obtained from DESI DR2 BAO + CMB (cosmic microwave background \citep{Planck:2018vyg,ACT:2023kun}), DESI DR2 BAO + CMB + Union3 SNIa (type Ia supernovae) \citep{Rubin:2023jdq}, DESI DR2 BAO + CMB + Pantheon+ \citep{Brout:2022vxf}, and DESI DR2 BAO + CMB + DES-Y5 (Dark Energy Survey Year 5 \citep{DES:2024jxu}), respectively (see Table VI in \citep{DESI:2025zgx}). In particular, the combination with DES-Y5 shows a tension $>4\sigma$, which is potentially significant.

In order to check consistency between DESI DR2 BAO and other datasets, especially DES-Y5 SNIa, we proposed in \citep{Dinda:2025hiu} a test of consistency between uncalibrated BAO and SNIa distances. We found that DES-Y5 data were in tension with DESI DR2 at $\sim\! 3\sigma$ for redshifts $z\sim1$, while the other SNIa datasets were consistent at $\lesssim 1\sigma$ for all $z$.

Recently, DES-Y5 has been updated to DES-Dovekie \citep{DES:2025sig} via an improvement of photometric cross-calibration between DES and low redshift surveys, retraining the SALT3 light curve model, and an improved approximation of host galaxy colour law. Note that this calibration is an internal calibration of SNIa data, which is different from the global calibration of absolute peak magnitude, $M_B$ of SNIa. In this analysis, by calibration, we refer to the second one, and our analysis is independent of the value of $M_B$. With this update, the deviation from $\Lambda$CDM reduces from $4.2\sigma$ to $3.2\sigma$ for DESI DR2 BAO + CMB + DES-Dovekie. We therefore expect that the inconsistency between DESI DR2 and DES-Dovekie has been reduced. In order to test this expectation, we re-do the analysis in \citep{Dinda:2025hiu} for the DES-Dovekie update --  confirming our expectation that there is no significant inconsistency amongst these datasets when they are compared using uncalibrated distances.

\section{Consistency test of uncalibrated BAO and SNIa distances}
\label{sec-relations}

The Alcock–Paczynski (AP) variable is
\begin{equation}
F_{\rm AP}(z) = \frac{\widetilde{D}_M(z)}{\widetilde{D}_H(z)} \, ,
\label{eq:FAP_bao}
\end{equation}
where $\widetilde{D}_M=D_M/r_d$ and $\widetilde{D}_H=D_H/r_d$ are the uncalibrated distances corresponding to  the comoving ($D_M$) and Hubble ($D_H$) distances and $r_d$ is the sound horizon at the drag epoch.

The distance modulus of SNIa is 
\begin{equation}
\mu_B(z) = m_B(z) - M_B = 5 \log_{10} \left[ \frac{d_L(z)}{{\rm Mpc}} \right] + 25  \quad \text{with} \quad d_L(z) = (1+z)D_M(z) \, .
\label{eq:2}
\end{equation}
Here $m_B$ and $M_B$ are the apparent and absolute peak magnitudes of SNIa.  Rearranging this equation we get
\begin{equation}
D_M(z) =
\begin{cases}
\alpha \, A(z) \, ,  & \quad \text{when } m_B \text{ is known} , \\
B(z) \, ,  & \quad \text{when } \mu_B \text{ is known} ,
\end{cases}
\label{eq:3}
\end{equation}
where
$b = {\ln(10)}/{5}$, and $\alpha = {\rm e}^{- b(5+M_B)} ~ {\rm Mpc}$,
and 
\begin{equation}
A(z) =(1+z)^{-1} { \exp \Big\{b\big[m_B(z)-20\big]\Big\}} \, , \quad B(z) = (1+z)^{-1}{ \exp \Big\{b\big[\mu_B(z)-40\big]\Big\}}  ~ {\rm Gpc} \, .
\label{eq:5}
\end{equation}
In the flat FLRW background, $D_H=D'_M$, where a prime is a redshift derivative, so that 
\begin{equation}
F_{\rm AP}(z) = \frac{D_M(z)}{D'_M(z)} \,
 =
\begin{cases}
{A(z)}/{A'(z)} \, ,  & \quad \text{when } m_B \text{ is known} , \\ \\
{B(z)}/{B'(z)} \, ,  & \quad \text{when } \mu_B \text{ is known} .
\end{cases}
\label{eq:FAP_sn}
\end{equation} 
DESI DR2 directly provides data for $F_{\rm AP}$. In addition, we use the  SNIa datasets Union3, Pantheon+, and DES-Dovekie. The first two have data up to $z\sim2.4$, whereas the third reaches $z\sim1.15$. We convert SNIa data to $B(z)$ data for Union3 and $A(z)$ data for the other two, using \autoref{eq:5}. We also compute the corresponding uncertainties in $A(z)$ or $B(z)$ using Gaussian propagation of uncertainty.

We apply a Gaussian Process (GP) \citep{Dinda:2022jih} to the $F_{\rm AP}$ data from DESI DR2 to reconstruct a smooth function and the associated uncertainties \citep{Gao:2025ozb,Gong:2025hoy}. Similarly, we apply GP to $A(z)$ or $B(z)$ data from SNIa, to reconstruct the function itself and its first derivative. We then apply Gaussian propagation of uncertainty to compute $F_{\rm AP}(z)$ from these functions. Note that we include all  cross-correlations to compute this uncertainty. We get all these cross-correlations automatically from a GP reconstruction. (For a review, see Appendix C of \citep{Dinda:2022jih}.) In this analysis, we assume a zero-mean function and a squared-exponential kernel covariance function. Note that, one can use other mean functions or kernel covariances, but the main results do not change significantly for these data (see discussions in Appendix F of \citep{Dinda:2024ktd} and Appendix B of \citep{Dinda:2025svh}).
\begin{figure}[!htbp]
\centering
\includegraphics[width=0.48\linewidth]{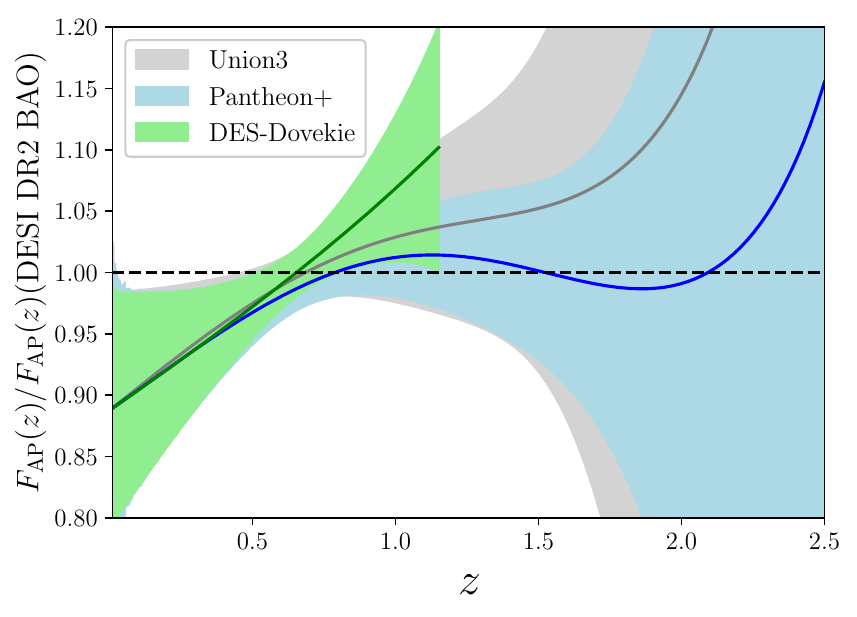}
\includegraphics[width=0.48\linewidth]{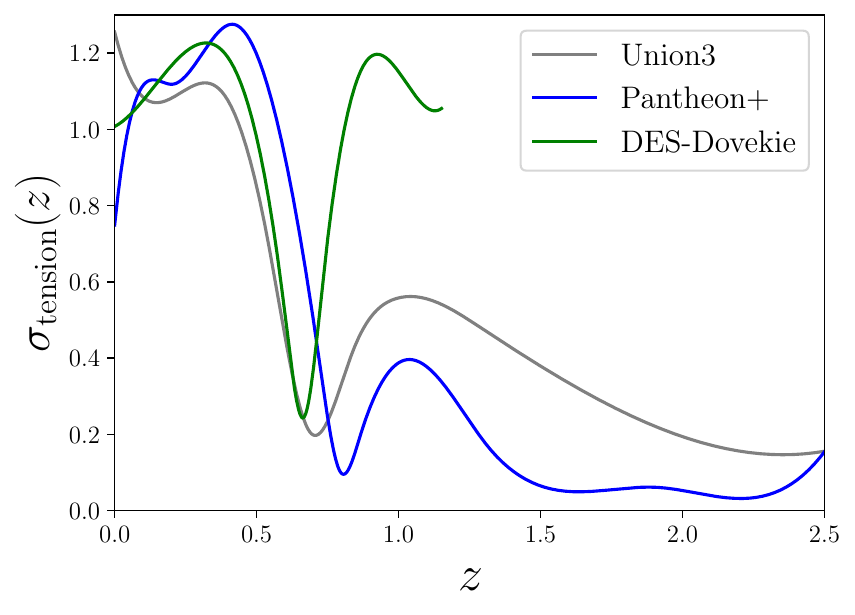}
\caption{{\em Left:} $F_{\rm AP}$ comparison of SNIa data to DESI data. {\em Right:} Tension between SNIa and DESI datasets, as given by \autoref{eq:tension}.
}
\label{fig:FAP}
\end{figure}
Using these GP predicted smooth functions, we compute $F_{\rm AP}(z)$ both from DESI DR2 and SNIa. We plot the ratios of reconstructed $F_{\rm AP}(z)$ from SNIa to the reconstructed $F_{\rm AP}(z)$ from DESI DR2 in the left panel of \autoref{fig:FAP}. We find that all these error regions overlap well within $\sim\! 1\sigma$. 

In order to properly quantify the consistency, we use a tension parameter in units of 1$\sigma$ confidence intervals \citep{Dinda:2025hiu}:
\begin{equation}
\sigma_{\rm tension}(z) = \frac{\big|F_{\rm AP}(z)-F_{\rm AP}(z)(\text{\small DESI DR2 BAO})\big|}{\Big\{\big[\Delta F_{\rm AP}(z)\big]^2+\big[\Delta F_{\rm AP}(z)(\text{\small DESI DR2 BAO})\big]^2\Big\}^{1/2}} \, .
\label{eq:tension}
\end{equation}
This tension parameter is shown in the right panel of \autoref{fig:FAP}. The maximum tension is $\sim\! 1.3\sigma$ around $z\sim0.5$ for Pantheon+. This means that the uncalibrated distances are consistent within $\sim\!1\sigma$ across DESI DR2 BAO and the three SNIa data sets. This result is independent of any systematics that arise solely from the $M_B$ calibration of SNIa data \citep{Dhawan:2024gqy}.

\section{Conclusions}
\label{sec-conclusion}

We tested the consistency between uncalibrated distance measures from DESI DR2 BAO and SNIa distance modulus measures from Union3, Pantheon+, and DES-Dovekie (updated from DES-Y5). Since BAO and SNIa measure different variables, we first convert them to a common variable, the Alcock–Paczynski ratio, which is ideally suited to test consistency.
By Gaussian Process reconstruction of $F_{\rm AP}$ and its predicted standard deviation, we found that the error regions about the means overlap with each other within $\sim\! 1\sigma$.
In order to quantify this consistency, we showed that the tension parameter for each SNIa dataset remains  $\lesssim 1\sigma$ across the redshift ranges. This confirms that the updated  DES-Dovekie dataset has removed the tension found by us in DES-Y5 \citep{Dinda:2025hiu}.

\acknowledgments
BRD and RM are supported by the South African Radio Astronomy Observatory and the National Research Foundation (Grant No. 75415). CC is supported by STFC grant ST/X000931/1.

\clearpage
\bibliographystyle{JHEP}
\bibliography{references}

\end{document}